\documentclass{article}

\usepackage{PRIMEarxiv}

\usepackage[utf8]{inputenc} 
\usepackage[T1]{fontenc}    
\usepackage{hyperref}       
\usepackage{url}            
\usepackage{booktabs}       
\usepackage{amsfonts}       
\usepackage{nicefrac}       
\usepackage{microtype}      
\usepackage{lipsum}
\usepackage{fancyhdr}       
\usepackage{graphicx}       
\graphicspath{{media/}}     
\usepackage{IEEEtrantools}
\usepackage{graphicx}
\usepackage{booktabs}
\usepackage{xcolor}
\usepackage{amsmath,amsfonts}
\usepackage{pifont}

\usepackage{tikz}

\newcommand{\bmu}{\boldsymbol{\mu}}

\newcommand{\bx}{\mathbf{x}}

\newcommand{\by}{\mathbf{y}}

\usepackage[title]{appendix}

\usepackage{float}
\usepackage{lineno}
\usepackage{amsmath}

\title{Probabilistic Wind Power Modelling via Heteroscedastic Non-Stationary Gaussian Processes

}

\author{
 Domniki Ladopoulou$^1$ \thanks{\textit{\underline{Corresponding author}}: 
\textbf{domna.ladopoulou.22@ucl.ac.uk}}$\quad$ Dat Minh Hong$^2$ $\quad$ Petros Dellaportas$^{1, 3}$\\
domna.ladopoulou.22@ucl.ac.uk\quad mdh58@cam.ac.uk\quad p.dellaportas@ucl.ac.uk\\
  $^1$ Department of Statistical Science, University College London, London, United Kingdom\\
  $^2$Department of Engineering, University of Cambridge, Cambridge, United Kingdom \\
  $^3$ Department of Statistics, Athens University of Economics and Business, Athens, Greece
}

\begin{document}
\maketitle

\begin{abstract}
Accurate probabilistic prediction of wind power is crucial for maintaining grid stability and facilitating the efficient integration of renewable energy sources. Gaussian process (GP) models offer a principled framework for quantifying uncertainty; however, conventional approaches typically rely on stationary kernels and homoscedastic noise assumptions, which are inadequate for modelling the inherently non-stationary and heteroscedastic nature of wind speed and power output. We propose a heteroscedastic non-stationary GP framework based on the generalised spectral mixture kernel, enabling the model to capture input-dependent correlations as well as input-dependent variability in wind speed--power data. We evaluate the proposed model on 10-minute supervisory control and data acquisition (SCADA) measurements and compare it against GP variants with stationary and non-stationary kernels, as well as commonly used non-GP probabilistic baselines. The results highlight the necessity of modelling both non-stationarity and heteroscedasticity in wind power prediction and demonstrate the practical value of flexible non-stationary GP models in operational SCADA settings.
\end{abstract}


\section{Introduction}

Wind power plays a pivotal role in the global transition toward renewable energy. By the end of 2024, the total installed wind power capacity worldwide had reached approximately 1,136 gigawatts (GW), representing an annual increase of 117 GW compared to 2023 \cite{gwec2025}. However, the variable and intermittent nature of wind generation introduces significant operational challenges to power systems. Therefore, accurate probabilistic wind power prediction is essential to ensure grid stability and efficient resource planning. In operational settings, most wind farms rely on supervisory control and data acquisition (SCADA) systems, which provide a cost-effective and widely deployed source of high-dimensional turbine measurements. These data offer substantial potential for accurate wind power prediction, but also introduce significant modelling challenges due to their peculiarities.

SCADA data collected from wind turbines are inherently noisy, high-dimensional, and shaped by complex operational variability. The relationship between wind speed and power output is nonlinear and regime-dependent, exhibiting two key forms of input-dependent variability: \emph{non-stationarity} and \emph{heteroscedasticity}. Non-stationarity arises when the correlation structure of the power curve changes across wind regimes (e.g., cut-in, rated, and cut-out), leading to locally varying smoothness and spectral content. Heteroscedasticity, in contrast, refers to input-dependent noise induced by sensor precision, control actions, and the temporal aggregation of high-frequency signals into 10-minute SCADA statistics. This aggregation amplifies regime-dependent variability: fluctuations are small under stable rated operation but large during transitional conditions, producing input-dependent noise levels.

A probabilistic formulation is therefore essential for characterising the full range of turbine operating conditions. By modelling the predictive distribution $p(y \mid \mathbf{x})$ rather than only its mean, one can quantify uncertainty, distinguish normal variability from abnormal behaviour, and support reliability assessment and decision-making under uncertainty. While heteroscedastic Gaussian process (GP) formulations have previously been applied to wind-power data to capture input-dependent variance~\cite{rogers2020}, the equally important problem of non-stationarity—where the covariance structure itself varies across operating regimes—remains underexplored in this context. Consequently, there is a need for models that can simultaneously capture these two fundamental sources of variability within a unified probabilistic framework.

The proposed heteroscedastic generalised spectral mixture (GSM) GP framework addresses these limitations by jointly modelling non-stationarity and heteroscedasticity within a single probabilistic formulation. By allowing both the kernel parameters and the noise variance to vary with the input, the model enables the predictive mean and uncertainty to adapt coherently across different turbine operating regimes, providing a flexible and interpretable representation of wind-turbine behaviour under uncertainty. Such adaptability is particularly valuable for real-world SCADA applications, where operating conditions can shift rapidly and uncertainty quantification is critical for operational decision-making.

Specifically, this study: (i) introduces a heteroscedastic non-stationary GP model based on the GSM kernel, capable of representing input-dependent correlations and observation noise;
(ii) demonstrates the benefits of this joint formulation through consistent improvements in both point-predictive and probabilistic metrics compared to stationary and singly-heteroscedastic GP baselines;
(iii) evaluates the framework under realistic SCADA constraints and pre-processing procedures, ensuring practical relevance for operational wind farm applications; and
(iv) provides empirical evidence that the proposed approach more accurately captures regime-dependent variability and uncertainty across different operating regions.

The remainder of this paper is structured as follows. Section~\ref{related_work} presents the related work. Section~\ref{methodology} describes the proposed heteroscedastic non-stationary GP framework. Section~\ref{experiments} outlines the experimental setup, including the data sources, preprocessing steps, training procedure, and evaluation metrics, and presents the corresponding results. Section~\ref{discussion} and ~\ref{conclusion} conclude with a discussion and a summary of the main findings.

\section{Related work}\label{related_work}
Recent reviews have highlighted the growing importance of hybrid models, probabilistic methods, and deep learning to advance wind power forecasting techniques \cite{review2023}. In particular, Lagos \textit{et al.} \cite{lagos2022} performed a literature review demonstrating the increasing scientific research toward probabilistic forecasting methods, hybrid modelling approaches, and uncertainty quantification techniques in the wind energy sector. We focus exclusively on probabilistic approaches, which generate full predictive distributions rather than point estimates.  This allows for uncertainty quantification, which is essential for risk-aware decision-making in power system operations.

GP methods have been extensively utilised for wind power prediction. Rogers \textit{et al.} \cite{rogers2020} proposed a heteroscedastic GP model that captures the input-dependent variance of wind turbine outputs, while Pandit \textit{et al.} \cite{pandit2020} incorporated turbine operational variables to enhance GP-based forecasts. Chen \textit{et al.} \cite{chen2014} integrated GP with numerical weather prediction outputs, achieving improved forecast accuracy. However, standard GPs rely on default stationary kernels, which are inadequate for capturing the complex dynamics of wind data. Wind speed and power output often exhibit pronounced non-stationary behaviour due to seasonal variations, atmospheric turbulence, and local weather dynamics. As such, stationary kernels are not appropriate to model these fluctuations effectively. To address challenges related to non-Gaussian uncertainties and time-varying characteristics, Kou \textit{et al.} \cite{kou2013sparse} proposed a sparse online warped GP model capable of adaptively learning non-Gaussian probabilistic distributions with reduced computational cost. Nevertheless, their approach primarily focuses on output warping and online adaptation, while maintaining a stationary covariance structure. 

Quantile regression-based methods have seen wide application. Yu \textit{et al.} \cite{yu2019probabilistic} proposed spatio-temporal quantile regression for regional wind power probabilistic forecasting. Later, Yu \textit{et al.} \cite{yu2021} introduced a deep quantile regression model to better capture non-linear dependencies. Zhou \textit{et al.} \cite{zhou2022} proposed composite conditional non-linear quantile regression to improve the accuracy of prediction intervals for very short-term regional wind power forecasting.  Other probabilistic modelling approaches have also emerged: Liao \textit{et al.} \cite{liao2022} applied a generative moment matching network to create realistic probabilistic scenarios without assuming specific distribution forms, while Zhang \textit{et al.} \cite{zhang2020} proposed an improved deep mixture density network to represent multimodal output distributions for wind power.

Several classical statistical approaches have also been introduced. Carpinone \textit{et al.} \cite{carpinone2015} adopted Markov chain models while  Dong \textit{et al.} \cite{dong2022arma} utilised multi-class autoregressive moving average modelling for wind power prediction. Ma \textit{et al.} \cite{ma2019} explored empirical dynamic modelling to reconstruct dynamic system behaviours directly from observed data.

Hybrid machine learning approaches are increasingly popular. Huang \textit{et al.} \cite{huang2023} proposed an a priori-guided and data-driven hybrid framework to combine physical insights with data-driven learning. Gu \textit{et al.} \cite{gu2023} proposed a hybrid framework combining fuzzy C-means clustering, whale optimisation algorithm-optimised extreme learning machine for wind power forecasting, and Gaussian mixture models for uncertainty quantification. Wang \textit{et al.} \cite{wang2022gmc} developed a hybrid deep neural network framework focused on improving probabilistic forecasts. Transfer learning for wind forecasting has been explored by Liu and Wang \cite{liu2022}, who applied multi-layer extreme learning machines to handle situations with limited training data. In addition, Fiocchi \textit{et al.} \cite{ff_dl_pd} proposed a probabilistic multilayer perceptron trained on SCADA data with heteroscedastic outputs and transfer learning across turbines for condition monitoring, demonstrating improved performance over other probabilistic models.

 Dong \textit{et al.} \cite{dong2021spatio} used spatio-temporal convolutional networks for forecasting the outputs of multiple wind farms. Krannichfeldt \textit{et al.} \cite{krannichfeldt2022} proposed an online ensemble approach for probabilistic wind power forecasting, capable of dynamically adapting to data changes. Eikeland \textit{et al.} \cite{eikeland2022} developed a deep learning-based probabilistic forecasting model suitable for Arctic regions characterised by complex topography. Zhang \textit{et al.} \cite{zhang2021mstn} proposed a multi-source temporal attention network that integrates heterogeneous numerical weather prediction data and historical measurements using temporal attention mechanisms to improve regional wind power probabilistic forecasting.

Che \textit{et al.} \cite{che2023} proposed a spatial-temporal probabilistic forecasting method based on multi-scale feature extraction and dynamic feature weighting, achieving improved performance in ultra-short-term wind power prediction settings. Meanwhile, Zhang \textit{et al.} \cite{zhang2023hybrid} developed a hybrid intelligent framework that integrates deterministic and probabilistic prediction approaches to enhance the accuracy and reliability of wind power prediction.
\section{Methodology}\label{methodology} 
We present a modelling framework guided by the need to provide a probabilistic and interpretable approach for wind power prediction. The proposed framework builds upon GP regression, which offers a principled Bayesian foundation for modelling uncertainty while maintaining analytical tractability. 
We first describe the standard GP regression model and discuss the limitations of stationary covariance kernels, which assume constant correlation structures across the input domain and are therefore inadequate for processes that exhibit regime-dependent dynamics. 
To address this, we introduce a non-stationary kernel formulation based on the GSM kernel, which allows the local adaptation of amplitudes, frequencies, and length-scales to capture smoothly varying operating regimes in wind turbine behaviour. 
Finally, we extend this formulation to a heteroscedastic setting, in which the observation noise variance itself depends on the input, enabling the model to represent input-dependent variability and uncertainty more accurately. 
Together, these components form a comprehensive heteroscedastic GSM–GP framework capable of learning both the non-stationary correlations and input-dependent noise structures inherent in wind power data, resulting in calibrated and physically interpretable uncertainty estimates across different operating conditions.

\subsection{GP regression}
In GP regression, we are given a data vector $\by =\{y_i\}_{i=1}^N \in \mathbb{R}^N$ whose entries are noisy evaluations of some function $f(\cdot)$ on a collection of $D$-dimensional vectors $X = \{\bx_i\}_{i=1}^N \in \mathbb{R}^{N \times D}$, i.e.~$y_i$ is a noisy observation of $f(\bx_i)$. In our setting, each input $\bx_i$ is a 10-minute wind speed measurement, and each output $y_i$ is the corresponding 10-minute active power.  We further assume that the noise $y_i-f(\bx_i)$ is independent Gaussian with mean $0$ and variance~$\sigma^2$. Moreover, we place a GP prior over $f(\cdot)$, with mean function $\mu(\cdot)$ and covariance kernel $k_\theta(\cdot,\cdot)$, so that the collection of function values $f(X):=[f(\bx_1),\ldots,f(\bx_N)]^T$ has a joint Gaussian distribution 
\begin{align*}
f(X) \sim \mathcal{N} (\mu(X),K(X,X)),
\end{align*}
where  $\mu(X) = [\mu(\bx_1),\ldots,\mu(\bx_N)]^T$, and $K(X,X)_{ij} = k_\theta(\bx_i,\bx_j)$, for all $i,j$. 
Setting  $A=K(X,X) + \sigma^2 I_N$, the log-marginal likelihood of the data becomes
\begin{align*}
\log p(\by|X) &= -\frac{1}{2} (\by-\mu(X))^T A^{-1} (\by-\mu(X))-\frac{1}{2} \log| A | -\frac{N}{2} \log(2\pi) \nonumber
\end{align*}
and the future observation $y^*$ with covariates $X^*$  have a normal conditional distribution with mean and variance given by
\begin{align} \label{mean}
E(y^* | \by) &= \mu(X^*)+K(X^*,X) A^{-1} (\by-\mu(X)), \\
V(y^* | \by) &= K(X^*,X^*) -K(X^*,X) A^{-1} K(X,X^*). \label{variance}
\end{align}

\paragraph{Stationarity in GP kernels.}
\label{kernels}
In GP regression, the kernel function encodes assumptions about the smoothness and similarity of the underlying function by defining how the correlation between outputs depends on the distance between input points. The most common choice in practice is to assume a stationary kernel, meaning that the covariance between two points depends only on their relative distance, not on their absolute locations in the input space. Formally, a kernel $k(\bx_i, \bx_j)$ is stationary if it can be expressed as a function of the lag vector $\boldsymbol{\tau} = \bx_i - \bx_j$, i.e., $k(\bx_i, \bx_j) = k(\boldsymbol{\tau})$. 
A typical stationary choice is the radial basis function (RBF) kernel, \begin{equation*}
    k_{\text{RBF}}(\mathbf{x}_i,\mathbf{x}_j | \theta) = \sigma_f^2 \exp\left(-\frac{1}{2} \sum_{d=1}^D \frac{(x_{i,d} - x_{j,d})^2}{\ell_d^2}\right),
\end{equation*} where $\theta=\{\sigma^2_f,\ell_1,\ell_2,\dots, \ell_D \}$ represents the kernel hyperparameters with $\ell_d$ denoting the length-scale parameter along dimension $d$ and $\sigma_f^2$ representing the output variance. Other stationary formulations, such as the Matérn family, share similar properties and are widely used due to their simplicity, analytic tractability, and ability to model smooth functions effectively.

The assumption of stationarity can be restrictive for real-world systems where the underlying signal exhibits regime changes or locally varying dynamics. In wind power generation, the relationship between wind speed and power output changes substantially across operating regions, leading to input-dependent smoothness and variance. Stationary kernels, which impose uniform behaviour across the input space, are unable to adapt to local changes in smoothness or signal variance. This necessitates the use of a non-stationary covariance function, which allows local adaptation of the kernel parameters and enables the model to represent heterogeneous dynamics more effectively.  
\subsection{GSM kernel}
\label{gsm}
   To model such input-dependent behaviour, we employ the GSM kernel \cite{Remes2017Non-StationaryKernels}, a non-stationary covariance function that allows input-dependent hyperparameters. The GSM kernel generalises the spectral representation of stationary kernels by introducing input-dependent amplitudes, frequencies, and length-scales, enabling flexible modelling of non-stationary correlations that evolve across the domain.
   
   Each mixture component $q$ in the GSM kernel is characterized by an input-dependent weight $w_q(x)$, frequency $\mu_q(x)$, and length-scale $\ell_q(x)$.  For notation convenience, we define $\ell_q(x)= v_q(x)^{-1/2}$ where $v_q(x)^{-1/2}$ denotes the local spectral variance. These functions are assigned independent GP priors with a zero mean function and a covariance matrix given by RBF kernels as follows \begin{align*}
\log w_q(x) &\sim \mathcal{GP}(0, k_w(x, x')), \\
\log \ell_q(x) &\sim \mathcal{GP}(0, k_{\ell}(x, x')), \\
\text{logit}\mu_q(x) &\sim \mathcal{GP}(0, k_{\mu}(x, x')).
\end{align*}
The resulting GSM kernel is given by
\begin{equation}
\label{GSM_kernel}
k_{\text{GSM}}(x_i, x_j) = \sum_{q=1}^{Q}  w_q(x_i)w_q(x_j) k_{\text{Gibbs},q}(x_i, x_j)  \cos\left(2\pi(\mu_q(x_i) x_i - \mu_q(x_j) x_j)\right),
\end{equation} where the Gibbs kernel \cite{GibbsKernel} is defined as
\begin{equation}
\label{eq:gibbs}
k_{\text{Gibbs},q}(x_i, x_j) = \sqrt{\frac{2\ell_q(x_i) \ell_q(x_j)}{\ell_q(x_i)^2 + \ell_q(x_j)^2}}
\exp \left\{-\frac{(x_i - x_j)^2}{\ell_q(x_i)^2 + \ell_q(x_j)^2}\right\}.
\end{equation} The Gibbs kernel acts as a non-stationary counterpart to the squared exponential kernel, incorporating input-dependent length-scales. While the above formulation is presented for one-dimensional inputs, it generalises naturally to multi-dimensional inputs $\bx_i, \bx_j \in \mathbb{R}^D$ through a product decomposition across dimensions
\begin{align*}
k_{\text{GSM}}(\bx_i, \bx_j | \theta) = \prod_{d=1}^{D} k_{\text{GSM}, d}(x_{i,d}, x_{j,d} | \theta_d),
\end{align*}
where  $\theta = (\theta_1, \dots, \theta_D)$  represents the input dependent hyperparameters for each mixture
$\theta_d = (w_{qd}, \ell_{qd}, \mu_{qd})_{i=1}^Q$ of the $N$-dimensional realizations $w_{qd}, \ell_{qd}, \mu_{qd} \in \mathbb{R}^N$ per dimension $d$. 

Due to the complex hierarchical structure of GPs with the GSM kernel, we can no longer perform maximum marginal likelihood inference directly. This is because computing the marginal likelihood requires integrating out not only the latent function  $f$  but also all input-dependent hyperparameter functions, making the integral intractable.  As an alternative, we employ maximum a posteriori (MAP) inference, where we maximise the log-posterior $\log p(\theta | \by) \propto \log p(\by|\theta) + \log p(\theta)$, where \( p(\by|\theta) \) is the marginal likelihood, integrating out only the function values  $f$. This allows us to perform gradient-based optimisation, using numerical optimisers such as ADAM.  

The final loss function is given by \begin{align*}
    \mathcal{L}(\theta) = \log \bigg( &\mathcal{N}_\by  \prod_{q,d=1}^{Q,D}  \mathcal{N}_{w_{qd}} \mathcal{N}_{\mu_{qd}} \mathcal{N}_{\ell_{qd}} \bigg),
\end{align*} where each component is

\begin{align*}
    \mathcal{N}_\by &= \mathcal{N}(\by \mid 0, K_{\theta} + \sigma^2 I) \\
    \mathcal{N}_{w_{qd}} &= \mathcal{N}(w_{qd} \mid 0, K_{w_d}) \\
    \mathcal{N}_{\mu_{qd}} &= \mathcal{N}(\mu_{qd} \mid 0, K_{\mu_d}) \\
    \mathcal{N}_{\ell_{qd}} &= \mathcal{N}(\ell_{qd} \mid 0, K_{\ell_d}). 
\end{align*}

The flexibility of the GSM kernel entails increased model complexity. Each mixture component defines several input-dependent functions—$w_q(x)$, $\mu_q(x)$, and $\ell_q(x)$—each governed by its own set of hyperparameters. As the number of mixture components $Q$ increases, the dimensionality of the parameter space expands, making optimisation more computationally demanding and increasing the risk of overfitting. Furthermore, the non-convex nature of the objective function renders the model sensitive to initialisation and prone to convergence to local optima. In practice, these challenges can be mitigated through multiple random initialisations and by experimenting with different mixture sizes $Q$ to improve robustness.

\subsection{Proposed heteroscedastic GSM--GP framework}

In standard stationary and non-stationary GP regression formulations, the observation noise is typically assumed to be homoscedastic, being characterised by a constant variance $\sigma^2$. However, many real-world processes exhibit input-dependent noise levels, a concept originally formulated in the GP framework by Goldberg \textit{et al.} \cite{goldberg1997regression}. In wind power generation, the variability of the observed power output is not constant across the operating range: it increases sharply within the mid-range wind-speed region (approximately 5–10 m/s) and decreases again near rated operation, where the turbine operates more stably. As shown in Fig.~\ref{fig:data}, this pattern appears as a distinct widening of the observation spread followed by a subsequent reduction, revealing pronounced input-dependent noise. This heteroscedastic behaviour motivates extending the GP framework to incorporate an input-dependent noise variance \cite{rogers2020}, modelled as a smooth function of the input, $\sigma^2(x)$ \cite{le2005heteroscedastic}.

To capture both input-dependent correlations and input-dependent noise structure, we propose a heteroscedastic GP model based on the GSM kernel introduced in Section~\ref{gsm}. 
The latent function $f(\cdot)$ is assigned a GP prior
\[
f \sim \mathcal{GP}\!\big(\mu(\cdot),\, K_{\text{GSM}}(\cdot,\cdot)\big),
\qquad 
\mu(x) = a + b x,
\]
where $\mu(x)$ is an affine mean function and $K_{\text{GSM}}(\cdot,\cdot)$ is the non-stationary covariance kernel defined in Eq.~\eqref{GSM_kernel}. 
Observations are modelled as
\begin{equation*}
y_i = f(x_i) + \varepsilon_i, 
\qquad 
\varepsilon_i \sim \mathcal{N}\!\big(0,\,\sigma^2(x_i)\big),
\end{equation*} where $\sigma^2(x)$ is a smooth, positive function representing the input-dependent noise variance. 
The overall covariance structure of the model is therefore
\begin{equation*}
K_{\text{HGSM}}(X,X)
= K_{\text{GSM}}(X,X) + \mathrm{diag}\!\big(\sigma^2(X)\big),
\end{equation*} where the first term models non-stationary latent correlations, and the diagonal term captures heteroscedastic uncertainty.

The input-dependent hyperparameters of the GSM kernel, $\{w_q(x), \ell_q(x), \mu_q(x)\}_{q=1}^Q$, together with the noise variance function $\sigma^2(x)$, are parametrised by compact neural networks that produce smooth, positive, and bounded outputs using sigmoid and softplus transformations. This ensures stable and interpretable variations across the input domain without requiring additional GP priors for these latent functions.

The model parameters $\theta$ consist of $a,b$ and the parameters of the neural networks that specify  $\{w_q(x), \ell_q(x), \mu_q(x)\}_{q=1}^Q$.  They
are learned by maximising the exact GP log-marginal likelihood, augmented with smoothness regularisation terms that penalise rapid changes in the log-length-scale and log-noise functions.  The loss function is
\begin{equation*}    
\mathcal{L}(\theta)
= -\log p(\by \mid X,\theta)
+ \lambda_{\sigma}\!\int\!\Big(\tfrac{d}{dx}\log \sigma^2(x)\Big)^{\!2}\!dx
+ \lambda_{\text{ns}}\!\sum_{q=1}^{Q}\!\int\!\Big(\tfrac{d}{dx}\log \ell_q(x)\Big)^{\!2}\!dx,
\end{equation*}
and is being estimated by 
\begin{equation*}    
\mathcal{L}(\theta)
= -\log p(\by \mid X,\theta)
+ \lambda_{\sigma} \sum_{i=1}^{N-1} \frac{ \left( \log \sigma^2(x_{i+1}) - \log \sigma^2(x_i) \right)^2 }
{x_{i+1} - x_i} 
+ \lambda_{\text{ns}}\!\sum_{q=1}^{Q} 
\sum_{i=1}^{N-1} 
\frac{ \left(
\log \ell_q(x_{i+1}) - \log \ell_q(x_i) \right)^2}
{x_{i+1} - x_i}
\end{equation*}
where $\lambda_{\sigma}$ and $\lambda_{\text{ns}}$ control the smoothness of the heteroscedastic and non-stationary components, respectively.  The regularisation encourages gradual spatial variation of the length–scale and noise functions, improving model stability and yielding more reliable uncertainty estimates under data sparsity.

Inference proceeds using the standard GP predictive equations, described in Eqs.~\eqref{mean}-\eqref{variance}, with the covariance matrix $K(X,X)$ replaced by $K_{\text{HGSM}}(X,X)$. 
This formulation retains the interpretability of the underlying kernel structure while ensuring statistically consistent and well-calibrated predictive distributions through exact GP inference. By allowing both the mean and variance to vary smoothly with wind speed, the model captures local changes in operating regimes. Model parameters are optimised via gradient-based MAP estimation with early stopping and multiple random initialisations to promote stable convergence and robustness to local minima. For completeness, stationary and non-stationary GP baselines without heteroscedasticity are also evaluated in Section~\ref{experiments} to assess the contribution of each modelling assumption.

\section{Experiments}
\label{experiments}
\subsection{SCADA data overview and filtering}

The study utilises SCADA and event data recorded at 10-minute intervals from a Senvion MM92 wind turbine located at Kelmarsh wind farm in the UK \cite{data}. The dataset spans from January $3$, $2016$, to July $1$, $2021$, comprising over $1.7$ million records across $110$ variables, including date-time, wind speed, and power output. In most cases, each data entry provides 10-minute averages, standard deviations, and the minimum and maximum values for all measured parameters. As wind speed summary statistics were first logged on September $25$, $2017$, only data from this date onward were included in the experiments.

An operational status and events file was utilised for data filtering to ensure consistent modelling. These files contain detailed information about the operational state of each wind turbine, including events such as technical failures, standbys, and warnings triggered by either operational or environmental factors. Data recorded during such events can influence the true relationship between wind speed and power output, leading to inaccurate model training. To mitigate this, periods corresponding to standbys, warnings, and operational stops were excluded. Additionally, data from the week leading up to each forced outage was removed to minimize the inclusion of unstable turbine behaviour before failure. A detailed illustration of the filtering process is provided in Section~4.1 of~\cite{ff_dl_pd}.

\begin{figure}[ht]
    \centering
    \includegraphics[width=\linewidth]{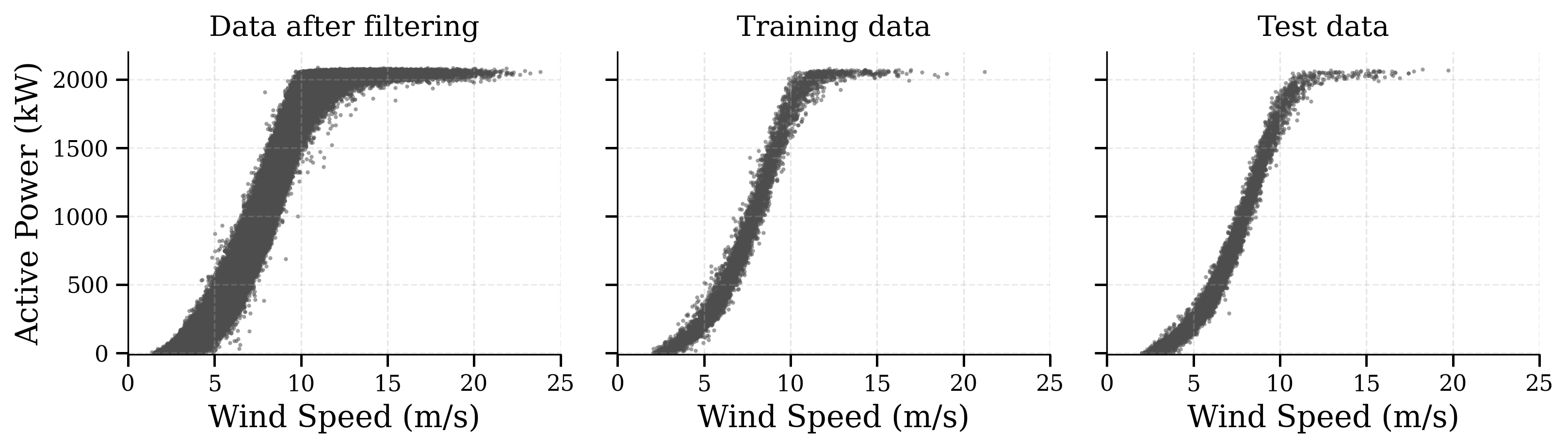}
\caption{Filtered SCADA dataset and data subsets used for model development and evaluation. 
The left panel shows the active power (kW) versus wind speed (m/s) after removing non-operational and faulted records. 
The middle and right panels display the training and test subsets, respectively, derived from temporally disjoint periods between 2017-2019 (training) and mid-2019 (testing). }
\label{fig:data}
\end{figure}

The impact of this filtering process is depicted in Fig.~\ref{fig:data}, where the filtered dataset, starting from September~25,~2017, is shown. 
Alongside the filtered power curve, we also indicate the subsets used for model development and evaluation. 
The training set comprises $7{,}000$ records collected between September~$25$,~$2017$ and June~$18$,~$2019$, whereas the test set contains $10{,}080$ records covering the period from June~$19$ to September~$5$,~$2019$, ensuring that model evaluation is conducted on an independent and temporally distinct subset of data.  To select the training set, the data were shuffled to ensure that observations from different seasons were included in the training set. For all analyses, wind speed (m/s) is used as the sole input variable and active power (kW) as the output target.  Wind speed and power are then standardised using the training-set statistics and predictions are de-standardised to physical units (m/s and kW) at test time. This normalisation ensures numerical stability between models.

\subsection{Experimental setup}

All models were trained on a single GPU. For each experiment, we reserved a set of $1{,}000$ additional wind–speed/power pairs as a validation set, sampled across the full wind regime and completely independent of both the training and test sets. This validation set was used exclusively for early stopping and model selection, with the stopping criterion based on the validation continuous ranked probability score (CRPS) defined in Section \ref{sec:metrics}.

The proposed heteroscedastic GSM–GP model was trained using the AdamW optimiser \cite{AdamW} with an initial learning rate of $10^{-2}$ and a weight decay of $5 \times 10^{-5}$. A preliminary warm-up GP with an RBF kernel was first trained to estimate the initial noise level $\sigma_0^2$, which was then used to initialise the bounds of the input-dependent noise network $\sigma^2(x)$. The number of spectral mixture components was fixed to $Q=3$ after evaluating alternative configurations with $Q \in \{2, 3, 4, 6\}$. Each kernel-parameter function—weights $w_q(x)$, frequencies $\mu_q(x)$, and length-scales $\ell_q(x)$—was parameterised by a compact neural network with a single hidden layer of $48$ $\tanh$-activated units, while the heteroscedastic noise function $\sigma^2(x)$ was modelled by a separate network with $32$ hidden units. Smoothness regularisation terms were applied to the logarithms of the noise and length-scale functions to enforce gradual spatial variation and improve generalisation.

To evaluate the performance of the proposed heteroscedastic GSM--GP, we compare it against a set of GP and non-GP probabilistic baselines representing different levels of flexibility and modelling assumptions. All models were trained and evaluated under the same normalisation and early-stopping criteria to ensure a fair comparison. A GP with an RBF kernel is included as a baseline for stationary GP performance.

To introduce greater flexibility, we also consider the spectral mixture (SM) kernel~\cite{Wilson2013GaussianExtrapolation} which generalises the RBF kernel by expressing the covariance as a weighted sum of Gaussian components in the spectral domain, following Bochner’s theorem~\cite{Bochner1959Lectures42}. For input lag $\boldsymbol{\tau} = \mathbf{x}_i - \mathbf{x}_j$, the kernel is defined as
\begin{equation*}
    k_{\text{SM}}(\boldsymbol{\tau} \mid \theta) = \sum_{q=1}^{Q} w_q \cos(2\pi \bmu_q^\top \boldsymbol{\tau}) \prod_{d=1}^{D} \exp(-2\pi^2 \tau_d^2 v_q^{(d)}),
\end{equation*}
where $Q$ is the number of mixture components in the spectral mixture, $w_q$ represents the relative contribution of the $q$-th component, $\bmu_q$ is a $D$-dimensional vector of spectral means, $\tau_d$ is the difference along the $d$-th dimension, and $v_q^{(d)}$ denotes the spectral variance of the $q$-th component along dimension $d$, controlling how rapidly correlations decay. In addition, $\bmu_q^{-1}$ corresponds to the periodicity of each component, while ${v_q^{(d)}}^{-1/2}$ represents the associated length-scale.

A key trade-off of the increased flexibility provided by the SM kernel is the larger number of hyperparameters. Specifically, each mixture component introduces three hyperparameters---$w_q$, $\bmu_q$, and $v_q^{(d)}$. Since the kernel is defined as a sum over $Q$ components, this results in a total of $3Q$ hyperparameters in our $1D$ setting, in addition to the noise variance. This high-dimensional parameter space increases the risk of overfitting and makes the optimisation process more sensitive to initialisation compared to standard kernels with fewer parameters. To mitigate these challenges, the GP model was trained with multiple random initialisations, and the configuration with the lowest training error was retained.

The SM GP model employed a constant mean function and a spectral mixture kernel with $Q=3$ components, selected consistently with the proposed heteroscedastic GSM--GP model. The mixture parameters were initialised using a distance-based rule: the inverse of the median spacing between input points was used to set the initial spectral means, and the empirical variance of the inputs was used to set the initial length-scales. Optimisation was performed by maximising the exact marginal log-likelihood using the Adam optimiser with gradient clipping, learning-rate scheduling, and early stopping.

The Gibbs kernel in Eq.~\eqref{eq:gibbs} is included as a non-stationary GP baseline. It can be viewed as a single-component special case of the GSM kernel, where only the local length-scale $\ell(x)$ varies with the input. The length-scale function is parameterised by a compact neural network with one hidden layer of 48 $\tanh$-activated units, bounded within $[0.05, 3]$ through a sigmoid transformation to ensure smooth and positive values. Training proceeds by maximising the exact GP marginal likelihood with an additional smoothness regularisation term on $\log \ell(x)$ to prevent rapid spatial fluctuations. Optimisation is carried out using the AdamW optimiser.

Two non-GP probabilistic baselines are also considered for comparison. A Bayesian neural network (BNN) is implemented following the Bayes-by-Backprop framework~\cite{blundell2015weight}, consisting of deterministic hidden layers and a mean-field Bayesian output layer. Each weight parameter is modelled as a Gaussian random variable with $\mathcal{N}(0, 0.3^2)$ priors. Other prior specifications did not alter the overall performance.  The network models a global homoscedastic observation noise $\sigma^2$, learned jointly with the weights. The training was based on the minimisation of the evidence lower bound (ELBO) using the AdamW optimiser.  Predictive means and variances are obtained through Monte Carlo sampling of the weight posterior at test samples.

Finally, the probabilistic method of bins extends the conventional deterministic ``method of bins'' commonly used in the wind industry to a fully probabilistic formulation. The wind-speed domain is discretised into bins, and within each bin, the empirical mean and variance of the power output are estimated with variance shrinkage to mitigate overfitting in bins with limited samples. These statistics are then interpolated smoothly across bins to yield continuous functions $\mu(x)$ and $\sigma^2(x)$, defining the conditional distribution $p(y \mid x) = \mathcal{N}\!\big(y \mid \mu(x), \sigma^2(x)\big).$

All GP models were trained using exact inference, which scales cubically with the number of training points, i.e.~$\mathcal{O}(N^3)$, due to the inversion of the covariance matrix. Consequently, the experiments were performed on a representative subset of $7{,}000$ training samples (see Fig.~\ref{fig:data}), selected to balance computational feasibility with adequate coverage of the wind-speed domain. Although sparse and variational GP formulations~\cite{titsias2009} can reduce complexity to $\mathcal{O}(N M^2)$ with $M \ll N$, exact inference was adopted here to isolate the effects of kernel design from those of approximate inference. This ensures that observed performance differences among models primarily reflect their representational capacity rather than inference approximations. The non-GP baselines, including the BNN and the probabilistic method of bins, scale linearly with the data size and thus offer complementary efficiency–accuracy trade-offs. Note that the probabilistic method of bins is applicable in this setting because only a single input feature is used. In contrast, all GP variants, as well as the BNN, could potentially benefit from the inclusion of additional input features.

\subsection{Evaluation metrics} \label{sec:metrics}

To assess both point-predictive accuracy and probabilistic calibration, we employ a diverse set of evaluation metrics that capture different aspects of predictive performance.  
For a vector of future observations $\{ x^*_i \}_{i=1}^{n}$ and $\{ y^*_i \}_{i=1}^{n}$ and corresponding generic predictive means 
$\{ \mu(x_i^*) \}_{i=1}^{n}$, the root mean square error (RMSE), mean absolute error (MAE) and 
the normalised mean absolute percentage error (NMAPE) 
are  defined, respectively, as
\begin{equation*}
\text{RMSE} = \left(\frac{1}{n} \sum_{i=1}^{n} (\mu(x_i^*) - y_i^*)^2\right)^{1/2}, 
\qquad 
\text{MAE} = \frac{1}{n} \sum_{i=1}^{n} |\mu(x_i^*) - y_i^*|,
\qquad 
\text{NMAPE}(\%) = \frac{100}{n \times C} \sum_{i=1}^{n} |\mu(x_i^*) - y_i^*|
\end{equation*}
where $C$ is the rated power of the wind turbine equal to $2050$~kW.  
Although these metrics offer valuable insights into how well our model predicts output power, they rely solely on the predicted means $\mu_i$ and are therefore beyond the primary focus of this work.

A metric to evaluate a model's probabilistic predictive performance is the negative log-predictive density (NLPD) which leverages the probabilistic nature of GPs.  The NLPD is calculated by taking the negative logarithm of the predictive probability density evaluated at the future observations averaged over all points $\{ y^*_i \}_{i=1}^{n}$.
Lower NLPD values indicate more appropriately calibrated predictive distributions.
The CRPS is a strictly proper scoring rule that measures the integrated squared difference between the predicted and empirical cumulative distribution functions~\cite{gneiting2007strictly}.  
For Gaussian predictive distributions, if $z_i$ are the normalised predictions, the CRPS can be expressed in closed form as
\begin{equation*}
\text{CRPS}
= \sigma_i \left[z_i (2\Phi(z_i) - 1) + 2\phi(z_i) - \frac{1}{\sqrt{\pi}}\right]
\qquad 
\end{equation*}
where $\Phi(\cdot)$ and $\phi(\cdot)$ denote the standard normal cdf and pdf, respectively.  
Lower CRPS values correspond to sharper and better-calibrated predictive distributions.
To evaluate the quality of predicted quantiles, we employ the quantile loss, which provides an asymmetric measure of accuracy for a desired quantile level $\tau \in (0,1)$:
\begin{equation*}
L_\tau(y_i, q_{\tau,i}) =
\begin{cases}
    \tau (y^*_i - q_{\tau,i}), & \text{if } y^*_i \ge q_{\tau,i}, \\[3pt]
    (\tau - 1)(y^*_i - q_{\tau,i}), & \text{otherwise,}
\end{cases}
\end{equation*}
where $q_{\tau,i} \equiv q_\tau(x^*_i)$ denotes the predicted $\tau$-quantile for input $x^*_i$.  
The loss is evaluated at $\tau = \{0.1, 0.5, 0.9\}$ to quantify predictive performance at different parts of the distribution.
The prediction interval quality is assessed using the Winkler score~\cite{gneiting2007strictly}, which jointly penalises interval width and miscoverage.  
For a nominal coverage level $1-\alpha$ with lower and upper interval bounds $[L_i, U_i]$, the Winkler score is defined as
\begin{equation*}
W_\alpha(y^*_i, L_i, U_i) =
(U_i - L_i)
+ \frac{2}{\alpha}(L_i - y^*_i)\mathbb{I}\{y^*_i < L_i\}
+ \frac{2}{\alpha}(y^*_i - U_i)\mathbb{I}\{y^*_i > U_i\},
\end{equation*}
where $\mathbb{I}\{\cdot\}$ is the indicator function.  
The score is reported for nominal levels of $80$\%, $90$\%, and $95$\%.  
Empirical coverage, defined as
$n^{-1} \sum_{i=1}^{N} 
\mathbb{I}\{L_i \le y^*_i \le U_i\}$,
is also computed to assess how closely the observed frequency of coverage matches the target level.
Finally, overall calibration across all reported intervals is summarised by the mean absolute coverage error (MACE), defined as
$
\text{MACE} = I^{-1}\sum_{i=1}^{I} \big| \hat{c}_i - c_i \big|,$
where $\hat{c}_i$ and $c_i$ denote the empirical and nominal coverages of the $i$-th interval, respectively. In our experiments, MACE is computed over the coverage probabilities corresponding to the nominal levels of $80$\%, $90$\%, and $95$\%. Lower MACE values indicate more consistent calibration across coverage levels.

\subsection{Results}

\begin{figure}[ht!]
    \centering
    \includegraphics[width=\linewidth]{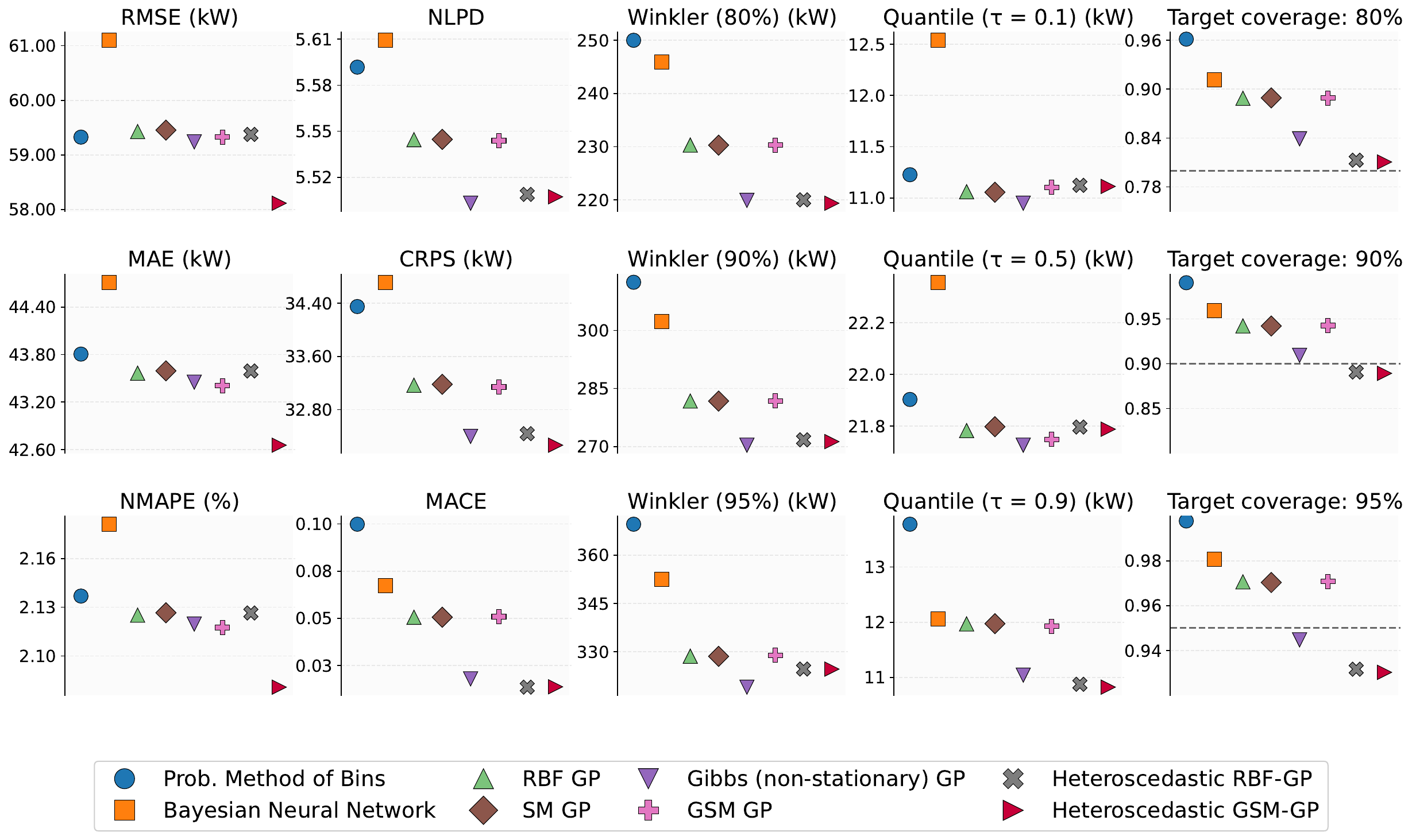}
    \caption{Comparison of point-predictive and probabilistic performance across all models. 
The first column reports point-error metrics (RMSE, MAE, NMAPE), and the second column presents probabilistic scores (NLPD, CRPS, MACE). 
The middle panel shows Winkler scores at nominal levels of $80$\%, $90$\%, and $95$\%. 
The fourth column displays quantile losses at $\tau = \{0.1, 0.5, 0.9\}$, while the last column shows empirical coverage at nominal levels of $80$\%, $90$\%, and $95$\%. 
Lower values indicate better performance across all metrics, whereas coverage closer to the nominal target (dashed dark grey line) reflects improved calibration.
    Abbreviations: CRPS~-- continuous ranked probability score, GP~-- Gaussian process, GSM~-- generalised spectral mixture, 
    MAE~-- mean absolute error, MACE~-- mean absolute coverage error, NMAPE~-- normalized mean absolute percentage error, 
    NLPD~-- negative log predictive density, Prob.~-- probabilistic, RBF~-- radial basis function, RMSE~-- root mean square error, 
    SM~-- spectral mixture.}
    \label{fig:metrics}
\end{figure}
The quantitative comparison across all evaluation metrics shown in Fig.~\ref{fig:metrics} demonstrates a clear progression in model performance as the assumptions of stationarity and homoscedasticity are gradually relaxed. Across all evaluation metrics, the GP models consistently outperform the non-GP baselines.
Both the BNN and the probabilistic method of bins exhibit inferior point-predictive accuracy and weaker probabilistic calibration.
Their RMSE, MAE, and NMAPE values are higher than those of the GP models, indicating less precise mean predictions.
Moreover, higher CRPS, NLPD, and Winkler scores demonstrate poorer uncertainty quantification, with both the BNN and the probabilistic method of bins producing overly wide, over-conservative intervals. 

In contrast, all GP variants provide more accurate and better-calibrated predictive distributions, reflecting the advantages of kernel-based formulations in capturing both correlation structure and predictive uncertainty.  Among the stationary GP baselines, the RBF and SM kernels achieve comparable levels of point-predictive accuracy, reflecting their ability to capture the overall nonlinear power–wind-speed relationship. However, both models exhibit limited probabilistic calibration, as evidenced by higher CRPS, Winkler scores, and coverage deviations, suggesting that their fixed covariance structures are less effective in capturing regions of varying variability.

Introducing non-stationarity into the covariance structure yields clear benefits in modelling the wind-speed–power relationship.
The non-stationary Gibbs GP, which allows the local length scale to vary with the input, achieves both higher point prediction accuracy and noticeably better probabilistic calibration than the stationary RBF and SM kernels. This demonstrates that accounting for regime-dependent smoothness is valuable for capturing the heterogeneous behaviours of the power curve.  Notably, the non-stationary Gibbs GP attains calibration performance comparable to the heteroscedastic GP variants and even surpasses them in the $95\%$ coverage probability.
The stationary GSM kernel also provides modest improvements in point-wise accuracy even without input-dependent noise. Overall, these results highlight the importance of incorporating non-stationarity when modelling the regime-dependent dynamics present in SCADA data.

Comparing the heteroscedastic GP variants highlights the added benefit of incorporating non-stationary latent correlations.
The heteroscedastic RBF-GP already improves upon all stationary counterparts, achieving lower NLPD, CRPS, and Winkler scores by capturing input-dependent noise levels that vary across the wind-speed domain. The proposed heteroscedastic GSM-GP extends this capability by introducing input-dependent kernel parameters in addition to input-dependent noise, thereby modelling both sources of non-stationarity jointly.
This dual flexibility yields the most accurate and well-calibrated predictive distributions overall, as evidenced by consistently lower CRPS values. The reduction in CRPS indicates that the predictive distributions are both sharper and better aligned with the observed variability, reflecting improved probabilistic calibration.

\begin{figure}[ht!]
    \centering
    \includegraphics[width=\linewidth]{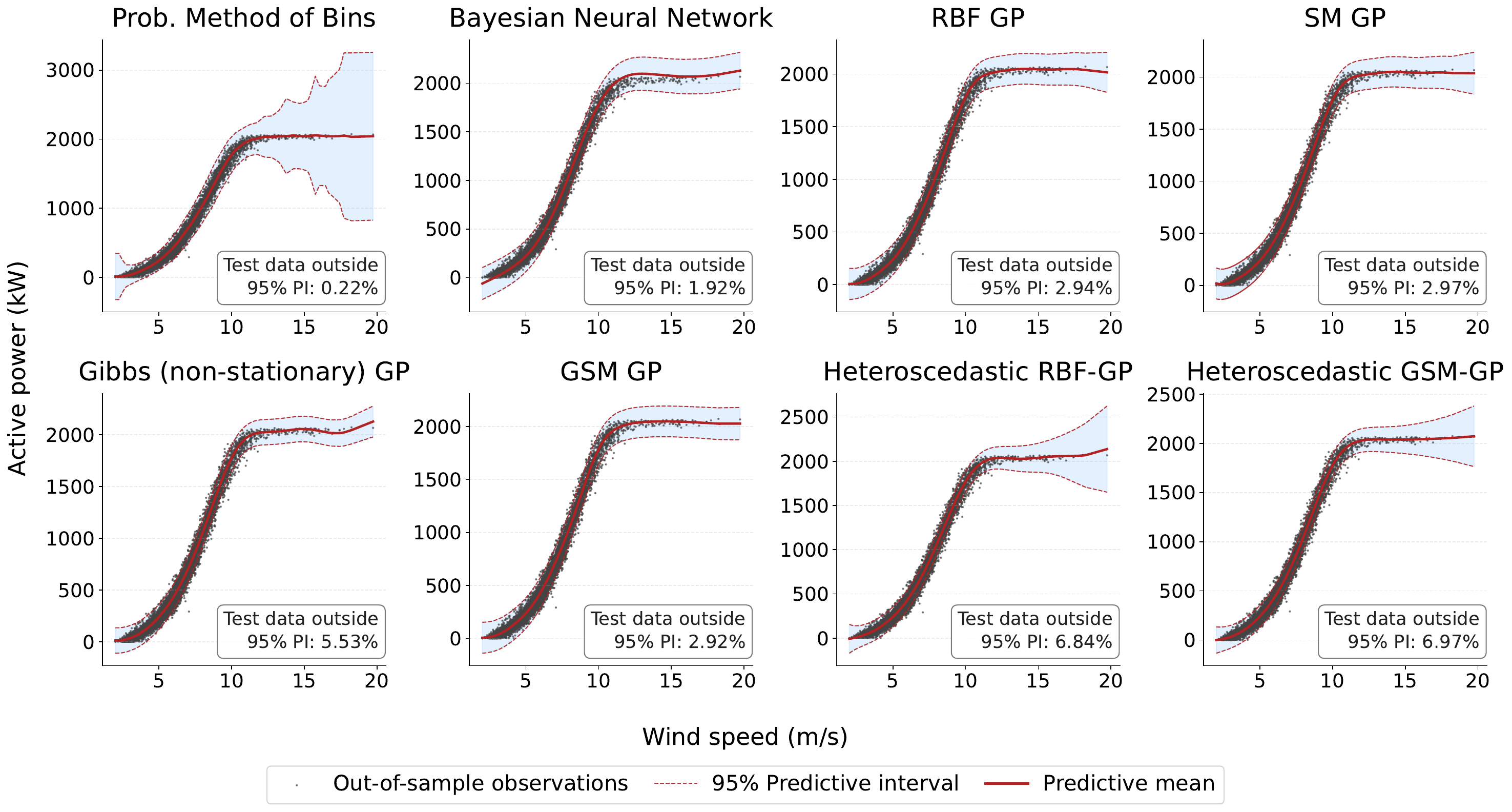}
    \caption{
    Predictive mean and 95\% predictive intervals (PI) for all models. 
    The shaded regions indicate the 95\% confidence intervals, and the percentage of test observations falling outside each interval is reported within each subplot. 
    Abbreviations: GP~-- Gaussian process, GSM~-- generalised spectral mixture, Prob.~-- probabilistic, RBF~-- radial basis function, SM~-- spectral mixture.
    }
    \label{fig:95PI}
\end{figure}

Fig.~\ref{fig:95PI} illustrates the predictive mean and corresponding 95\% predictive intervals, revealing consistent trends with the quantitative metrics.  As expected, the non-stationary Gibbs and the heteroscedastic GP models exhibit the most reliable uncertainty representation.  The heteroscedastic GPs show narrow confidence bounds in regions densely covered by training data and wider intervals where data are sparse.
This pattern is particularly evident near the turbine’s rated operating point, where the sample density declines sharply and the model appropriately expands its predictive variance to reflect increased uncertainty.

\section{Discussion}
\label{discussion}
Since GP posteriors are conditionally Gaussian, the resulting predictive distributions are inherently symmetric around their mean.
In contrast, the empirical power output near the nominal wind-speed region exhibits pronounced skewness, driven by aerodynamic saturation and turbine control dynamics.
Since the Gaussian likelihood models only the first two moments—the mean and variance—this asymmetry cannot be fully represented, leading to a mild underestimation of tail probabilities and a slight under-representation of extreme power values.  Despite this limitation, the proposed heteroscedastic GSM–GP on average outperforms both the rest of the GP variants and the non-GP baselines.

While recent advances in deep learning, such as long short-term memory networks, have achieved impressive results in sequence-modelling tasks, these approaches typically require large training datasets and extensive hyperparameter tuning to produce stable probabilistic behaviour \cite{ff_dl_pd}. In contrast, GP models provide a more transparent and data-efficient alternative, yielding well-defined predictive distributions that remain reliable even in relatively small datasets—an aspect particularly valuable for newly installed wind farms where limited operational data are available for wind power modelling. The kernel formulation of GPs allows direct interpretation of smoothness, variability, and local non-stationarity in the underlying relationship between wind speed and power, and model inference is less sensitive to random initialisation or overparameterisation. Although deep neural architectures can represent complex temporal dependencies, the proposed non-stationary GP framework remains competitive in terms of probabilistic calibration, interpretability, and performance in limited-data regimes.

The computational complexity of GP models constrains their scalability to very large datasets. This limitation could be alleviated in future work through sparse or variational formulations with inducing points \cite{titsias2009}. Similarly, hierarchical and multi-output GP designs can help manage the growth in kernel hyperparameters introduced by additional input features, though at increased modelling effort. Extensions that model the noise variance as a separate GP \cite{lazaro2011variational} enable heteroscedastic inference within a variational framework, allowing input-dependent uncertainty to be captured without increasing the overall computational complexity relative to standard GP formulations. Overall, while the current study focuses on exact inference to isolate kernel effects, these scalable and heteroscedastic formulations offer a path for extending the proposed approach to larger datasets, where it would remain competitive with alternative probabilistic deep-learning models.

\section{Conclusion}

This work demonstrated that wind turbine power curves exhibit strong non-stationarity and input-dependent noise, making heteroscedastic non-stationary modelling essential for accurate uncertainty quantification. We introduced a heteroscedastic GSM–GP, in which all kernel parameters and the observation noise variance are learned as smooth input-dependent functions. Unlike classical stationary GPs or sparse variational approaches, our model performs exact inference on a representative subset of SCADA data and focuses on improving the expressiveness of the kernel itself.

Across all experiments, the proposed model achieved improved probabilistic performance while maintaining competitive accuracy in mean predictions. The comparison against multiple baselines, including RBF GPs, SM GPs, Gibbs GPs, BNNs, and a probabilistic method of bins, highlights that heteroscedastic non-stationary power curve modelling is crucial for faithfully capturing the dispersion patterns observed in real wind turbine behaviour.

The resulting probabilistic representation provides a foundation for downstream tasks such as condition monitoring, anomaly detection, and energy forecasting, where accurate uncertainty estimates directly influence decision quality. Future work may explore extensions to higher-dimensional feature spaces, sparse or variational approximations for larger datasets, and integration with physical or hybrid models for improved extrapolation.

\section{Author contribution statement}
D.L. conceptualised the study, developed the methodology for the heteroscedastic non-stationary
Gaussian Process framework, conducted all model experiments, validated the results, carried out the
formal analysis, curated the data, prepared the visualisations, and wrote and revised the manuscript.
D.M.H. performed the experiments for the RBF, SM, and GSM baselines and contributed to writing the first
draft of the manuscript. P.D. supervised the entire project.
\label{conclusion}

\bibliographystyle{unsrt}  
\bibliography{references}

\end{document}